\documentclass[english]{article}
\usepackage[latin9]{inputenc}
\usepackage{textcomp}
\usepackage{amstext}
\usepackage{graphicx}

\makeatletter

\DeclareFontEncoding{LGR}{}{}
\DeclareRobustCommand{\greektext}{%
  \fontencoding{LGR}\selectfont\def\encodingdefault{LGR}}
\DeclareRobustCommand{\textgreek}[1]{\leavevmode{\greektext #1}}
\ProvideTextCommand{\~}{LGR}[1]{\char126#1}

\newcommand{\lyxaddress}[1]{
\par {\raggedright #1
\vspace{1.4em}
\noindent\par}
}

\makeatother

\usepackage{babel}
\begin{document}

\title{``How to squash a mathematical tomato'', Rubic's cube-like surfaces
and their connection to reversible computation.}

\author{{\Large{}Ioannis Tamvakis}}
\maketitle

\lyxaddress{\begin{center}
Sainsbury lab, Cambridge University, Cambridge, United Kingdom
\par\end{center}}
\begin{abstract}
Here we show how reversible computation processes, like Margolus diffusion\cite{margolus1988physics},
can be envisioned as physical turning operations on a 2-dimensional
rigid surface that is cut by a regular pattern of intersecting circles.
We then briefly explore the design-space of these patterns, and report
on the discovery of an interesting fractal subdivision of space by
iterative circle packings. We devise two different ways for creating
this fractal, both showing interesting properties, some resembling
properties of the dragon curve. The patterns presented here can have
interesting applications to the engineering of modular, kinetic, active
surfaces. 
\end{abstract}

\section*{Introduction}

In cellular automata theory, reversible computation processes have
been devised on a regular grid of cells to simulate, in a discrete
fashion, diffusion \cite{margolus1988physics,margolus1998crystalline}
or systems capable of universal computation\cite{fredkin2002conservative,margolus1988physics}.
Some of these processes take the form of alternating the application
of the automaton rules in 2 distinct subdivisions of the square grid
of the automaton (figure 1a). In each timestep, the update rules are
designed to be reminiscent of an exchange of values between neighbouring
positions of the grid. By doing this, the number of positions with
the same state does not change, and conservation of simulated particles
is warranted. An interesting analogy to this procedure that we can
think of is to cut a physical surface by two identical square packings
of circles, the second one shifted in respect to the first one so
that the circles of these two packings intersect perpendicularly (figure
1b). This allows, by independent rotation of the circles of either
one or the other packing, for specific degrees (multiples of the right
angle), the shuffling of the lenses to different positions. If we
assign boolean values to the lenses, The act of shuffling them by
rotation using the same rules as the automaton we are trying to recapitulate
can give the correct evolution of the system. This discovery made
us wonder how such a system can be physically realised. A surface
can be cut out this way and the pieces can possibly be designed as
to allow for rotation with minimal interlocking due to misalignment.
This can give a modular, kinetic tiling where a specific lens can
move to any other lens position, not unlike how the pieces of a Rubik\textquoteright s
cube move. We have yet to find an interesting engineering approach
to make these pieces mobile in a regular fashion. Another question
that arises is what other patterns using circles can lead to a well
structured grid of positions that could recapitulate known cellular
automata. A candidate pattern is to start with a hexagonal circle
packing and then increase (figure 1c), or double the radius of the
circles (figure 1d). In the last tiling (figure 1d), essentially every
position of the surface is kinetic.

\begin{figure}[h]
\centering{}\includegraphics[width=10cm]{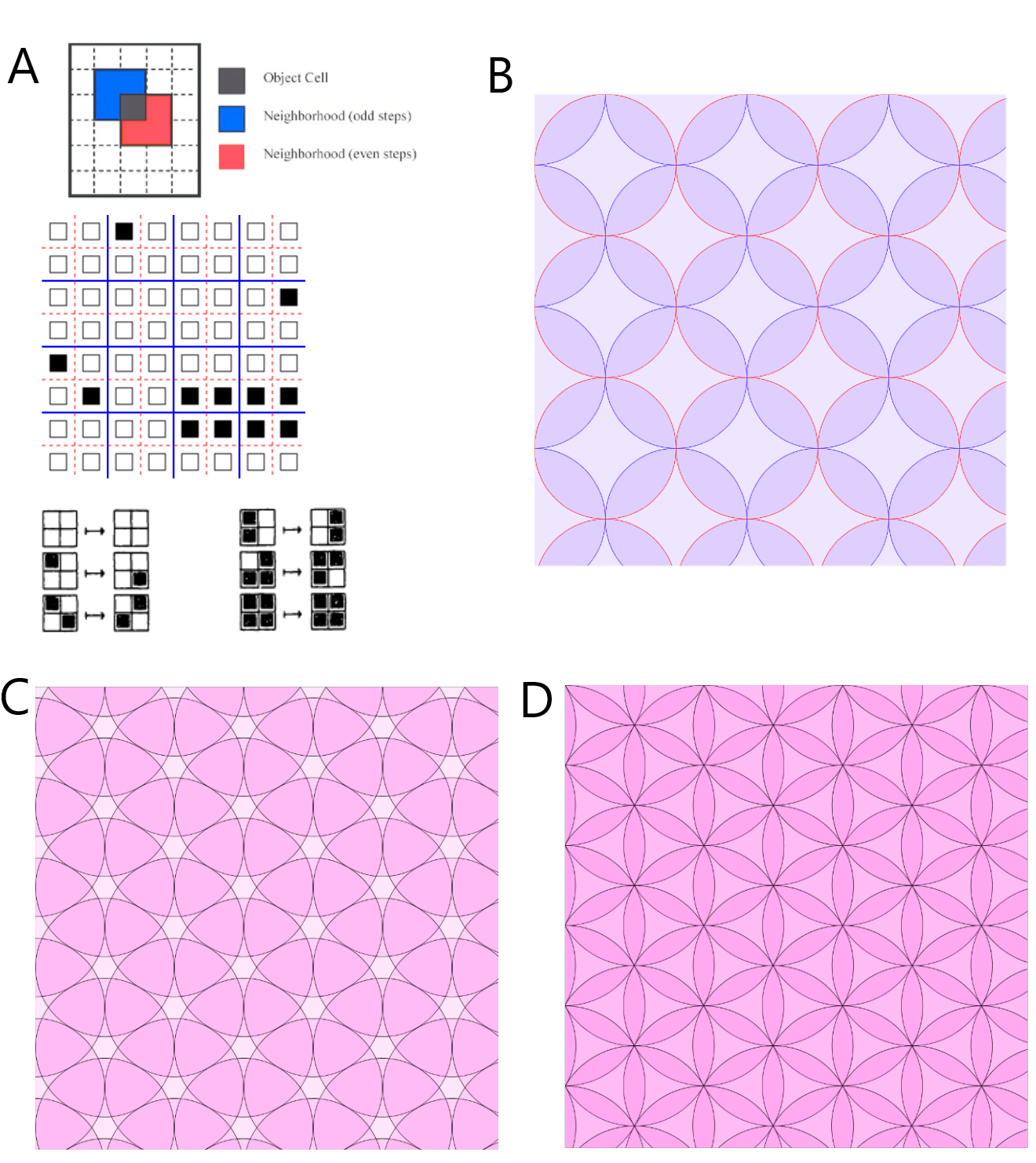}\caption{From cellular automata to the shuffling of pieces in regular circle
packings.\textbf{ A: }In the Margolus diffusion block cellular automaton,
the grid is partitioned in blocks of four, and each timestep the subdivision
is shifted by diagonaly. The update rules of the cellular automaton(bottom,
addopted from\cite{margolus1988physics}) are applied. Notice how
in all cases the change in values can be envisioned as a rotation
of the 4-block.\textbf{ B:} Analogous system made by two square circle
packings (red, blue). If we cut the surface in this way, canonical
rotation of the circles lead to the free movement of the lenses. The
rules of the block automaton can be applied. \textbf{C: }Starting
from one triangular circle packing, by extending the radii, we can
make a modular, kinetic surface where the main pieces move on a honeycomb
grid. \textbf{D: }By cutting the surface in this way, every position
is free to move about using rotations of the circles. }
\end{figure}

We were further interested in how we can produce fully kinetic tiled
surfaces, so we explored a bit more the design space without considering
its possible connection to automata theory. As the circles do not
have to be the same size, and smaller circles could give an arbitrarily
better precision of movement to tiles, we strived to find a natural
geometric construction using circle packings that in theory could
move, by rotating the circles, any position of the divided space to
any other. We found the following natural fractal construction to
be interesting.

\subsection*{First fractal definition}

The fractal \emph{\textquotedblleft T\textquotedblright{}} can be
made by taking an infinite circle packing in 2D space (circles of
the same radius, touching each other, with their centers defining
a isosceles triangular lattice), and applying to it, in each iteration
of the fractal, the following transformation: $\left[\begin{array}{c}
x'\\
y'
\end{array}\right]=\left[\begin{array}{c}
\sqrt{7/3}\\
\sqrt{7/3}
\end{array}\right]\left[\begin{array}{cc}
cos(\pi/6) & sin(\pi/6)\\
-sin(\pi/6) & cos(\pi/6)
\end{array}\right]\left[\begin{array}{c}
x\\
y
\end{array}\right]$ This simply states that we take the starting circle packing $N_{x}$,
and we make a new one $N_{x+1}$ by turning it by 30 degrees, and
scaling it by $tan(30\text{º)}$. This leads to a second, smaller
circle packing, with few of the smaller circles having the same center
as the big ones and all the others being at the intersection of any
three of the larger ones (figure 2a-d). This transformation can be
applied iteratively on each new circle packing, to get the fractal
\emph{\textgreek{T} }with any number of levels (or its inverse to
get the $N_{x-1}$ level) . Figure 2e represents the fractal after
infinite iterations. We observed that the space is not homogeneously
subdivided, so canonical rotations might lead to ``jamming'' of
the circles of distant fractal levels. After 3 iterations of the fractal,
some circles are not radialy symmetric anymore, in terms of inner
pieces arrangement. 

\begin{figure}[ph]
\centering{}\includegraphics[width=12cm]{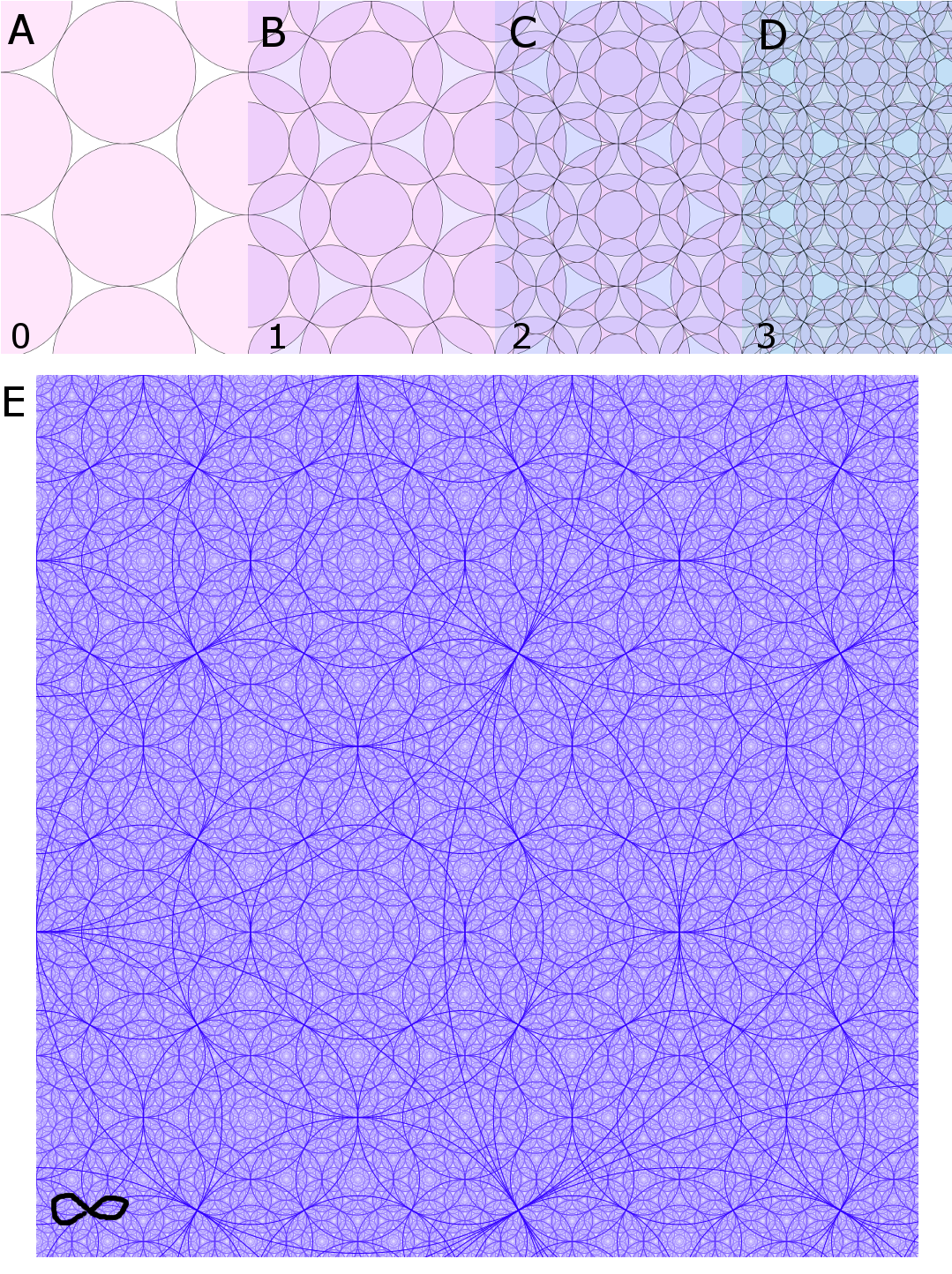}\caption{\textbf{A-D}: The fractal T after applying the production rule 0-3
times. \textbf{E: }The fractal T after infinite iterations. The transparency
of the circle coloring is increased at every level. Visualisation
were made using Processing\cite{fry2001processing} }
\end{figure}

\subsection*{Second fractal definition}

The fractal\emph{ T} can also be made by applying a production rule
to each circle of the original circle packing separately. Each cirlce
must have a vector denoting its polarity. The polarity vectors for
the circles of the original circle packing are all the same, pointing
all in the same direction of one of the angle bisectors of the triangular
lattice defined by their centers. We can then define the following
fractal construction \emph{\textgreek{W}} for each of the circles.
For each circle of radius $r_{n}$, polarity angle $p_{n}$, we create
three new circles of radius $r_{n+1}=r_{n}\cdot tan(\pi/6)$, touching
externally serially, their centers on the line defined by the polarity
angle $p_{n}$. The center of the middle circle has the same center
as the original bigger circle, and one new circle is above and one
bellow. The polarity of the new circles is $p_{n+1}=p_{n}+\pi/6$
(they are rotated +30º degrees after placement) (figure 3a). We can
then apply the same procedure on the new circles to get multiple iterations
of the fractal \emph{\textgreek{W}}, getting in each iteration a curve
which we would like to name \emph{Athena\textquoteright s curve} (\emph{``A}'')\cite{mara2013social},
reminiscent of the \emph{dragon curve}\cite{davis1970number} (figure
3c). Like the latter, \emph{A} is able to tile the 2D space, in any
iteration, to give a complete circle packing (figure 3b). The sum
over the iterations of Athena\textquoteright s curve (the fractal
\emph{\textgreek{W}}), when applied to all circles of the original
circle packing \emph{N}, gives the fractal $T$. An interesting property
of Athena\textquoteright s curve is that the total area that the circles
occupy, $E(A_{x})$, is invariable to the iteration number $x$, as
$E(A_{x})=3^{x}\cdot\pi\cdot(r\cdot(tan(\pi/6))^{x})^{2}=\pi\cdot r^{2}$,
where $r$ is the original circle's radius, and the produced circles
do not overlap. The length of the curve $L(A_{x})$ however diverges
to infinity as $L(A_{x})=3^{x}\cdot2\cdot\pi\cdot r\cdot(tan(\pi/6))^{x}=2\cdot\pi\cdot3^{(x/2)}$.
An intriguing open problem is to find what is the area the fractal
\emph{\textgreek{W}}, in its combined infinite iterations, encompasses. 

\begin{figure}[ph]
\centering{}\includegraphics[width=12cm]{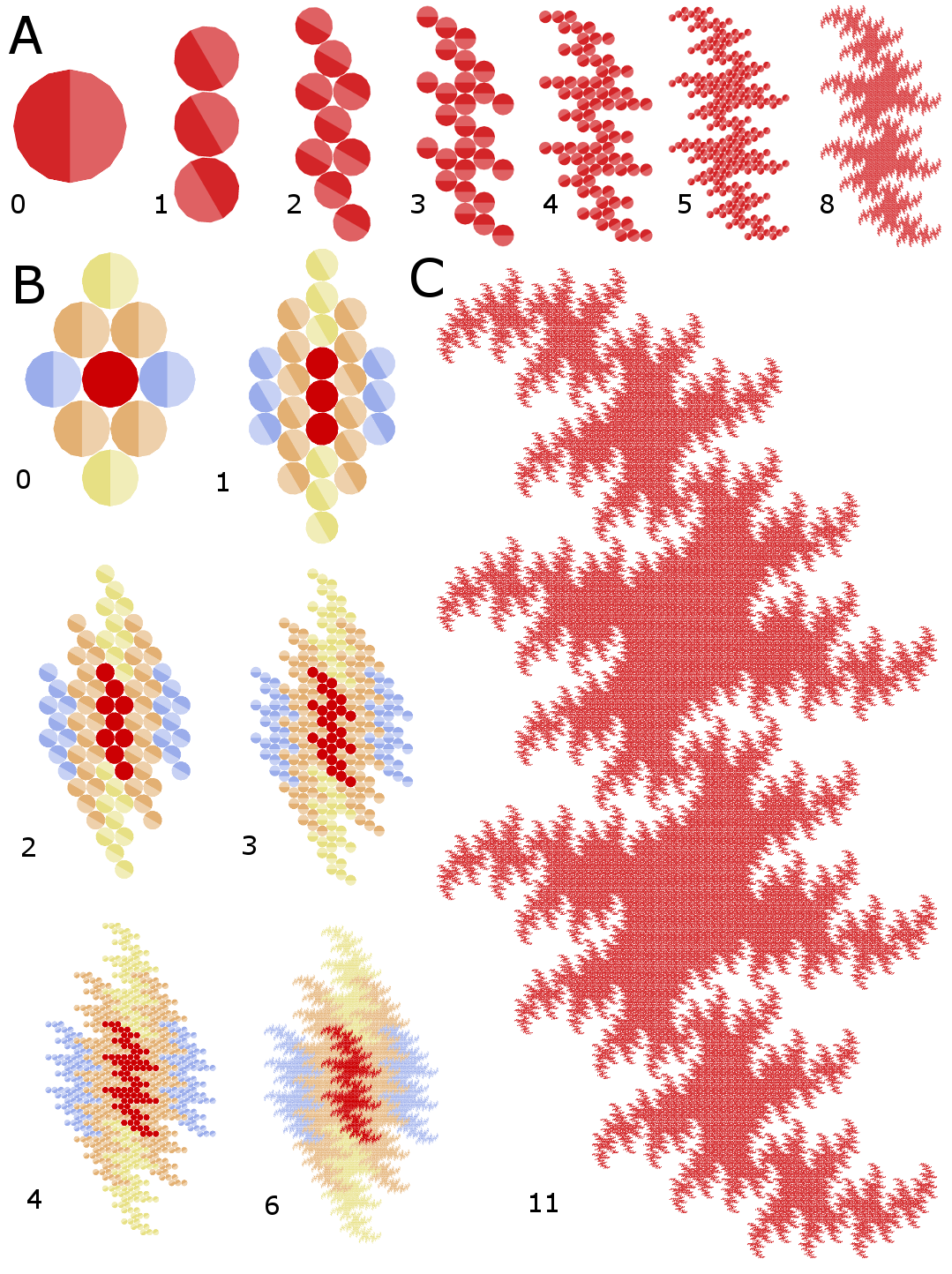}\caption{``How to squash a mathematical tomato''\textbf{ A:} A circle becomes
an Athena's curve, after applying the production rule the denoted
times. The area of the resulting circles stays the same. Two different
colors at each side of the circles serve to denote polarity axes.\textbf{
B: }Athena's curves can tile the space, interlocking perfectly with
eachother. \textbf{C: }Athena's curve after 11 applications of the
production rule. Images were made using Lpy\cite{boudon2010py}}
\end{figure}

\section*{Discussion}

Here we present a mapping, from discrete reversible computational
processes, made in cellular automata, to the shuffling of pieces created
by cutting a two-dimensional surface by overlaping circle packings.
We find this connection amusing, as it means that we can simulate
block automatons in a physical setting. This surface geometry might
have interesting applications. For example one can envision a surface
cut in this manner to be kinetic, its pieces being free to move about,
mobilised by a robotic understructure, without sacrificing the surface
integrity. Pieces that need to be replaced can then move to the periphery,
and changed there. Pieces that have specific properties, on the other
hand, can be directed to the part of the surface that are needed most.
At the same time, block automaton rules can be applied and the pieces
will recapitulate, with their movement, biophysical processes. 

We strived to find a recursive subdivision of space, using this general
formula of overlaying circle packings, that would be able to move,
by the rotation of the circles, the pieces formed, effortlessly. This
is an interesting open problem, namely, if it is feasible to construct
a fractal subdivision of space, using circles, that keeps the radial
symmetry of the pieces inside each circle.\\

\textbf{Aknowledgements:} I would like to thank Eugenio Azpeitia for
helping in creating the second fractal definition, using Lpy, and
for interesting discussions.

\bibliographystyle{plain}
\bibliography{xampl}

\end{document}